\documentclass{jpp}
\usepackage{graphicx}
\usepackage{booktabs}

\usepackage[utf8]{inputenc}
\usepackage[T1]{fontenc}
\usepackage{amsmath}

\shorttitle{Bootstrap Current Modeling}
\shortauthor{S. Saxena, N.M. Ferraro, M.F. Martin, A.M. Wright}

\title{Bootstrap Current Modeling in M3D-C1}

\author{Saurabh Saxena\aff{1}
  \corresp{\email{ssaxena@pppl.gov}},
  Nathaniel Ferraro\aff{1},
  Mike F. Martin\aff{2},
  Adelle M. Wright\aff{3}}

\affiliation{\aff{1}Princeton Plasma Physics Laboratory, Princeton, NJ, USA \aff{2}Thea Energy, Kearny, NJ, 07032, USA \aff{3}University of Wisconsin--Madison, Madison, WI, USA

}

\begin{document}

\maketitle

\begin{abstract}
Bootstrap current plays a crucial role in the equilibrium of magnetically confined plasmas, particularly in quasisymmetric (QS) stellarators and in tokamaks, where it can represent bulk of the electric current density. Accurate modeling of this current is essential for understanding the magnetohydrodynamic (MHD) equilibrium and stability of these configurations. This study expands the modeling capabilities of M3D-C1, an extended-MHD code, by implementing self-consistent physics models for bootstrap current. It employs two analytical frameworks: a generalized Sauter model (\cite{sauter1999neoclassical}), and a revised Sauter-like model (\cite{redl2021new}). The isomorphism described by \cite{landreman2022optimization} is employed to apply these models to quasisymmetric stellarators.   The implementation in M3D-C1 is benchmarked against neoclassical codes, including NEO, XGCa, and SFINCS, showing excellent agreement. These improvements allow M3D-C1 to self-consistently calculate the neoclassical contributions to plasma current in axisymmetric and quasisymmetric configurations, providing a more accurate representation of the plasma behavior in these configurations.  A workflow for evaluating the neoclassical transport using SFINCS with arbitrary toroidal equilibria calculated using M3D-C1 is also presented.  This workflow enables a quantitative evaluation of the error in the Sauter-like model in cases that deviate from axi- or quasi-symmetry (\textit{e.g.}, through the development of an MHD instability).
\end{abstract}

\section{Introduction}
Bootstrap current is a neoclassical electrical current in the plasma driven by inhomogeneities in the magnetic field and is proportional to temperature and density gradients in the plasma (\cite{peeters2000bootstrap}, \cite{helander2012classical}).  In tokamaks and quasisymmetric stellarators, the bootstrap current can represent a significant fraction of the total current density, and it can strongly affect the rotational transform of the magnetic field (\cite{neuner2021measurements,helander2011bootstrap}). Therefore, accurate calculation of this current is essential for capturing corrections due to neoclassical physics.

Neoclassical transport can be calculated by solving the drift-kinetic equation. While several codes exist to solve this equation, obtaining a self-consistent state requires solving both the drift-kinetic equation and the MHD equilibrium equations iteratively (\cite{landreman2022optimization}).This process is computationally intensive (\cite{redl2021new}). As a more efficient alternative, analytical models can be employed to approximate neoclassical transport. One such model is the approach taken by \cite{sauter1999neoclassical}, which provides a local expression for the bootstrap current and neoclassical conductivity. In this model, the coefficients relating the bootstrap current to temperature and pressure gradients are obtained by fitting results from the codes CQL3D (\cite{killeen2012computational}, \cite{monticello1993summary}) and CQLP (\cite{sauter19943}) applied to a range of axisymmetric equilibria. This model, however, is known to be less accurate at higher electron collisionalities $\nu_e^*>1$ limiting its applicability near the plasma edge (\cite{landreman2012local}, \cite{koh2012bootstrap}). The model relies on three neoclassical parameters, fraction of trapped particles $f_t$, collisionality $\nu^*$, and effective charge number $Z_{eff}$. To overcome these limitations, a revised version of the Sauter model, developed by \cite{redl2021new}, utilizes the NEO code (\cite{belli2008kinetic}, \cite{belli2011full}), a drift-kinetic solver for neoclassical steady-state solutions. The revised model offers enhanced accuracy and robustness across a broader range of collisionalities, extending its applicability beyond the limitations of the original Sauter model. Both the Sauter model and the Redl model have been extensively verified and tested in tokamak geometry.

However, these models are not generally applicable to stellarators because they are exclusively fit to calculations in axisymmetric geometry. A recent approach by \cite{landreman2022optimization} addresses this by exploiting the isomorphism between axisymmetric and quasi-symmetric configurations. This method allows the application of the Redl model to compute the bootstrap current in quasi-symmetric stellarators. Here, quasisymmetry refers to the special condition in which the magnetic field strength $B=|B|$ exhibits continuous symmetry in a suitable coordinate system. Specifically, $B$ depends only on the the flux surface label ($\psi$) and a linear combination of Boozer poloidal ($\theta$) and toroidal angles ($\zeta$), such that $B=B(\psi,\theta-N\zeta)$ (\cite{landreman2022optimization}). This symmetry ensures conserved guiding-center motion analogous to that in axisymmetric fields (\cite{landreman2019quasisymmetry}).

While this development offers a significant step forward in applying analytical neoclassical models to stellarators, challenges remain in integrating these advances into comprehensive simulation frameworks. Although several MHD codes have been developed for tokamak applications, there is still a notable gap in the ability of nonlinear MHD simulation codes to treat stellarators. Most MHD codes assume axisymmetric computational domains and are designed for tokamak applications and are only able to handle weakly-shaped stellarator geometries ((\cite{schlutt2012numerical}, \cite{schlutt2013self}, \cite{roberds2016simulations}). Recent advancements, however, have enabled M3D-C1 to model the nonlinear MHD evolution of strongly-shaped stellarator plasmas by accommodating non-axisymmetric domain geometries (\cite{zhou2021approach}).  The ability to treat stellarator geometry has also recently been implemented in JOREK3D (\cite{Nikulsin_2022}) and NIMSTELL (\cite{Sovinec_2021}).

In this work, we further extend the capabilities of the M3D-C1 code to include self-consistent physics models for bootstrap current for both tokamak and quasi-symmetric stellarator geometry. For the calculation of bootstrap current, we use the \cite{sauter1999neoclassical} formula and its improved version described in \cite{redl2021new}. Building on the method developed by \cite{landreman2022optimization}, we apply isomorphism between axisymmetric and quasi-symmetric geometries to compute the bootstrap current for quasisymmetric stellarator configurations.

The remainder of the paper is organized as follows. Section~\ref{sec:model} provides an overview of the M3D-C1 code, including implementation details of the two bootstrap current models. This section also outlines the neoclassical models used for validation and verification. Section~\ref{sec:results} details the computational setup and presents results of cross-verification between M3D-C1 and the neoclassical models for a tokamak case and two quasiaxissymteric stellarator cases. Additionally, this section presents simulations of the nonlinear evolution of a QA stellarator equilibrium, highlighting the impact of the bootstrap model. Finally, Section~\ref{sec:conclusions} offers a summary of the findings and a discussion of their implications.

\section{Model Description}
\label{sec:model}

M3D-C1 is a high-fidelity extended-MHD code (\cite{jardin2012multiple}). It employs a split-implicit time scheme, which allows for time steps that extend beyond the Alfv\'enic timescale, enabling stable simulations on the transport timescale (\cite{jardin2012review}). The code utilizes high-order finite elements with $C^1$ continuity, constructed on an axisymmetric mesh. Recent developments have enabled M3D-C1 to model the MHD evolution of stellarator plasmas by accommodating non-axisymmetric domain geometries (\cite{zhou2021approach}). The governing equations for both tokamak and stellarator simulations are fully detailed in \cite{jardin2012multiple} and \cite{zhou2021approach}, while the single-fluid model equations, relevant to the analysis of plasma behavior in this work are reproduced in Appendix~\ref{appC}.

In the present work, we extend M3D-C1 to include a non-inductive current source, specifically, the bootstrap current, by modifying Ohm's law as follows:
\begin{equation}
\boldsymbol{E}=-\boldsymbol{v}\times\boldsymbol{B}+\eta[\boldsymbol{J}-\boldsymbol{J}_x].
\end{equation}
This is equivalent to adding a force on the electrons $\boldsymbol{F}_x^{e}$ such that $\boldsymbol{F}_x^{e}=-\eta n_e e\boldsymbol{J}_x=-\boldsymbol{R}_u$. Where, $-\boldsymbol{R}_u$ is the friction force on the electrons due to collisions with ions (\cite{braginskii1965transport}). Thus, $\boldsymbol{J}_x$ is the current that would arise from balancing the force on the electrons $\boldsymbol{F}_x^{e}$ with the force from collisions with ions. Even if the force is applied instantly, the actual current in the plasma $\boldsymbol{J} = \boldsymbol{\nabla} \times \boldsymbol{B} / \mu_0$ will evolve on
resistive timescales. How this force appears in the ion momentum equation depends on whether this is an internal force (\textit{e.g.}, the bootstrap current) or an external force. For an internal force, there must be an equal and opposite force in the ion momentum equation, which results in no net force added to the MHD force balance equation.

For a force consistent with a purely parallel, divergence-free current ($\boldsymbol{J}_x=J_{\parallel} \frac{\boldsymbol{B}}{B}$), the quantity $J_{\parallel}/B$ must remain constant on a magnetic surface. Using this condition, we can relate the local parallel current to the magnetic field via the expression:
\begin{equation}
J_{\parallel}=\frac{\langle \boldsymbol{J}_x \cdot \boldsymbol{B}\rangle}{\langle B^2\rangle}\boldsymbol{B},
\end{equation}
where $\langle \cdot \rangle$ denotes the magnetic surface average. To calculate $\langle \boldsymbol{J_x \cdot B} \rangle$, the generalized neoclassical models of \cite{sauter1999neoclassical} and \cite{redl2021new} are employed, which provide analytical expressions for this quantity.  The specific forms of these models are given in Appendix~\ref{appA}. 

For stellarators, the bootstrap current is computed using the method introduced by \cite{landreman2022optimization} (shown in Eq.~\ref{eq:Landreman}), which exploits the isomorphism between axisymmetric and quasi-symmetric geometries: 

\begin{align}
\langle \boldsymbol{J_x \cdot B}\rangle 
&= \frac{\tilde{G} }{\iota - N} \Bigg(
    L_{31} \left[ 
        p_e \frac{\partial \ln n_e}{\partial \psi_t} 
        + p_i \frac{\partial \ln n_i}{\partial \psi_t} 
    \right] \notag \\
&\quad + p_e (L_{31} + L_{32}) \frac{\partial \ln T_e}{\partial \psi_t}
    + p_i (L_{31} + \alpha L_{34}) \frac{\partial \ln T_i}{\partial \psi_t}
\Bigg)
\label{eq:Landreman}
\end{align}

Here, $\tilde{G} (\psi_{t})=G+NI$, $G$ is $\mu_0/2\pi$ times the poloidal current outside the flux surface $(\psi_{t})$, $I$ is $\mu_0/2\pi$ the toroidal current inside the flux surface $(\psi_{t})$, $\psi_{t}$ is toroidal flux per radian i.e. $\Psi_t = 2\pi\,\psi_t$, $\iota$ is the rotational transform, $p_{e/i}, n_{e/i}, T_{e/i}$ are the electron (ion) pressures, densities and temperatures, and $\alpha$, $L_{31}$, $L_{32}$, $L_{34}$ are the bootstrap coefficients (see Appendix~\ref{appB} for further definitions). The Sauter-Redl-Landreman formulation requires information about the global equilibrium, quantities such as $G,I,$ and $ \psi_t$.  This information is generally not known within M3D-C1, as its treatment of magnetic field does not assume the presence of magnetic surfaces.  To address this challenge, a separate calculation is performed in which an approximate magnetic coordinate system is constructed from M3D-C1 output using Fusion-IO (\cite{ferraro_fusionio}), taking isotherms of $T_e$ as proxies for magnetic surfaces.  Due to the strongly anisotropic conduction of $T_e$, these isotherms closely coincide with magnetic surfaces when surfaces exist and dynamics are sufficiently slow; when surfaces do not exist, the isotherms are related to quadratic flux minimizing surfaces (\cite{Dewar94}).  However, it is worth reiterating that the Sauter-Redl-Landreman formula is valid only in quasisymmetric geometries.  Limitations introduced by deviations from quasi-symmetry are discussed in section~\ref{sec:stellarator_verification}.

To evaluate the neoclassical bootstrap current density and its related coefficients correctly, it is necessary to compute quantities such as the trapped particle fraction $f_t$ and geometric factors like the inverse aspect ratio $\epsilon$ and $qR$ (as defined in Appendix~\ref{appB}). These parameters are integral to the expressions for the bootstrap current coefficients $\alpha$, $L_{31}$, $L_{32}$, and $L_{34}$ (see \cite{redl2021new} for detailed expressions). To facilitate these calculations, we define a coordinate system based on the electron temperature isotherms, $\hat{T}_e=1-T_e / T_e^{max}$ with $T_e^{max}$ being the maximum electron temperature within the plasma domain. This is roughly analogous to the normalized flux $\Psi_{t_N}=\psi_t/\psi_{LCFS}$. Within this framework, the global equilibrium quantities $I(\hat{T_e})$, $G(\hat{T_e})$, $\iota(\hat{T_e})$ along with $f_t(\hat{T_e})$, $\epsilon(\hat{T_e})$, $qR(\hat{T_e})$, and $\iota(\hat{T_e}) = (d\Psi_p/d\hat{T_e}) / (d\Psi_t/d\hat{T_e})$ are calculated externally to M3D-C1. They are then read into M3D-C1 at the start of the simulation. During the dynamical simulations, the bootstrap coefficients $\alpha$, $L_{31}$, $L_{32}$, and $L_{34}$ are evaluated locally at each time step using the evolving profiles of $\hat{T}_e$. Using $\hat{T}_e$ rather than $T_e$ enables a consistent treatment of cases where the temperature profile evolves in amplitude but maintains its shape, thereby avoiding unnecessary recomputation of the global quantities and other variables required for the calculation of the bootstrap coefficients.

To validate the accuracy of the bootstrap current calculations in M3D-C1 simulations, the results are compared with predictions from well-established neoclassical codes, namely: 
(1) XGCa, a global total-f gyrokinetic neoclassical code, a detailed description of which can be found in \cite{hager2016gyrokinetic} 
(2) NEO, a drift-kinetic neoclassical steady-state solver, a detailed description of which can be found in (\cite{belli2008kinetic,belli2011full})
(3) stellarator neoclassical code SFINCS (the Stellarator Fokker-Planck Iterative Neoclassical Conservative Solver), which solves the radially local 4D drift-kinetic equation without assuming quasisymmetry.  See \cite{landreman2014comparison} for further details. 
These comparisons are performed for both tokamak (NEO, XGCa, SFINCS) and quasi-symmetric
stellarator  (SFINCS) geometries, ensuring consistent and reliable results across a range of magnetic confinement configurations.

\section{Numerical Results: Bootstrap Current Calculations}
\label{sec:results}
\subsection{Tokamak Case Verification}
\label{sec:tokamak_verification}

\begin{figure}
  \centering
  \includegraphics{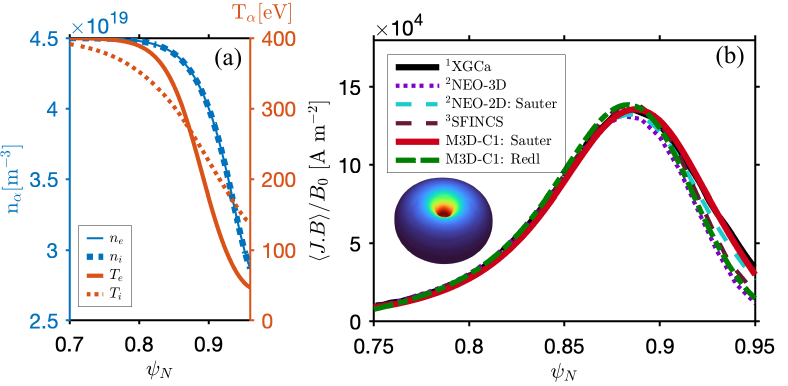}
  \caption{Setup and results for tokamak verification:(a) prescribed density and temperature distributions; (b) bootstrap current profiles showing that M3D-C1's results closely approximate those from the drift-kinetic calculations.$^1$\cite{hager2016gyrokinetic},$^2$\cite{belli2008kinetic,belli2011full},$^3$\cite{landreman2014comparison}.}
\label{fig:1}
\end{figure}
For the verification study, the low aspect ratio ``CIRC1'' case from \cite{hager2016gyrokinetic} is considered here. The configuration features a circular cross-section with inverse aspect ratio $\epsilon = 0.84$ at the outer boundary. In this analysis, following \cite{hager2016gyrokinetic}, the inverse aspect ratio $\epsilon$ is defined as the ratio of the mean minor radius $a_{mean} = (R_{max} - R_{min})/2$ to the geometrical center $R_c=
(R_{max} + R_{min})/2$ of a flux-surface, where $R_{max}$ and $R_{min}$ are the
maximum and minimum major radii on a flux surface. The density and temperature profiles used in the simulations are tanh-type pedestals as shown in Fig. \ref{fig:1}a. 

Figure \ref{fig:1}b shows the bootstrap current profile for the CIRC1 configuration. Maximum bootstrap current occurs at $\psi_N \sim 0.89$, where $\epsilon= 0.63$ and the electron collisionality $\nu_e^*$ is 0.92. A comparison between the M3D-C1 code using the \cite{sauter1999neoclassical} model and the NEO code’s Sauter model reveals a 2.18\% difference at the peak of the bootstrap current profile. Furthermore, the bootstrap current profiles from M3D-C1 and XGCa are nearly identical, with a difference of only 0.04$\%$ at the peak. The M3D-C1 profile calculated using the \cite{redl2021new} model shows a 2.12\% difference from XGCa and a 1.01$\%$ difference from SFINCS profiles at the peak.
It is important to highlight that the electron collisionality in this configuration is less than unity ($\nu_e^* < 1$), which is within the regime where the Sauter and Redl models are expected to provide reasonable approximations. While no single code serves as a definitive reference, NEO and SFINCS solve the drift-kinetic equation directly and provide high-fidelity neoclassical transport calculations. Given the complexity of bootstrap current calculations, agreement within a few percent at the profile peak is generally considered acceptable. Accordingly, the strong agreement among M3D-C1, NEO-3D, SFINCS, and XGCa simulations supports the validity of the bootstrap current calculations for this case.

\subsection{Stellarator Verification}
\label{sec:stellarator_verification}
\begin{figure}
  \centering
  \includegraphics[width=\textwidth]{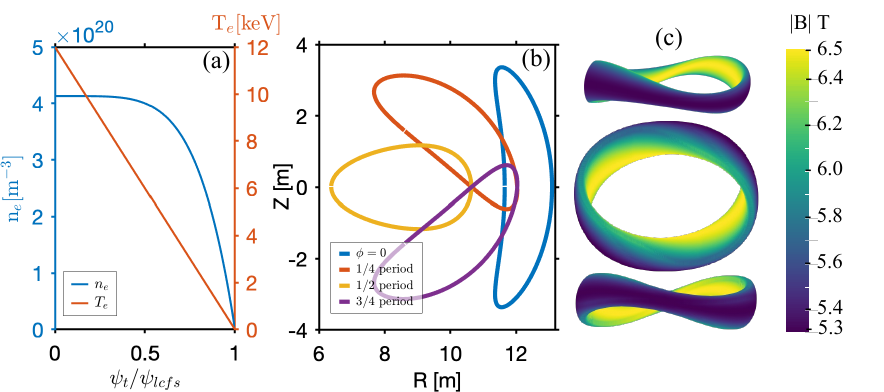}
  \caption{Quasi-axissymetric configuration (\texttt{QA\_Case1}) (a) Density and temperature equilibrium profiles, (b) toroidal cross-sections of the plasma boundary, and (c) three-dimensional view.}
\label{fig:2}
\end{figure}

\begin{figure}
  \centering
  \includegraphics[width=\textwidth]{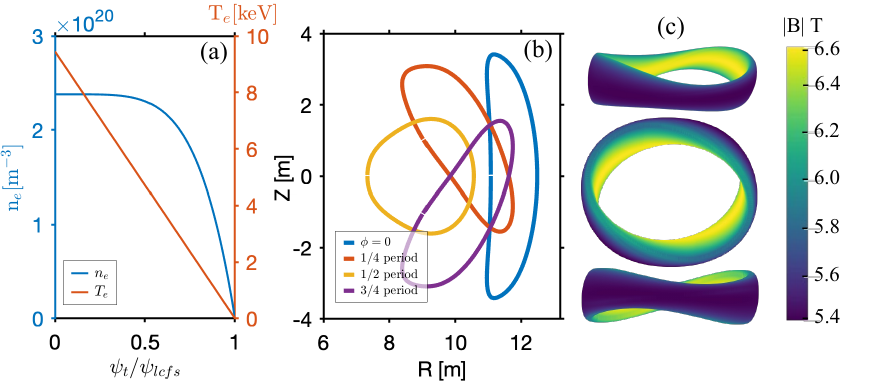}
  \caption{Optimized QA configuration with a volume-averaged $\beta=$2.5$\%$ (\texttt{QA\_Case2}) (a) Density and temperature equilibrium profiles, (b) toroidal cross-sections of the plasma boundary, and (c) three-dimensional view.}
\label{fig:3}
\end{figure}

The implementation of \cite{redl2021new} formulae with the isomorphism as defined by \cite{landreman2022optimization} are verified on two quasi-axisymmetric (QA) stellarator configurations from \cite{landreman2022magnetic} and \cite{landreman2022optimization}. These QA configurations both have a minor radius of 1.70 m and volume-averaged ${{B}}=5.86$ T. The density ($n$) and temperature ($T$) profiles for both cases are specified using:
\begin{subequations}
\begin{equation}
n(\psi_{t_N})=n_0(1-\psi_{t_N}^5),
\end{equation}
\begin{equation}
T(\psi_{t_N})=T_0(1-\psi_{t_N})
\end{equation}
\end{subequations}
where $\psi_{t_N}$ is the normalized toroidal flux. For the first case (\texttt{QA\_Case1}), a pure hydrogen plasma with $n_0=n_{H,0}=4.13 \times 10^{20} $ m$^{-3}$ and$T_0=T_{H,0}=12$ keV (\cite{landreman2022optimization}, \cite{landreman2022optimization} - Sect. IV) is considered. This case uses a VMEC equilibrium with zero plasma pressure. The bootstrap current is calculated using the temperature and density profiles shown in Fig. ~\ref{fig:2}a,, however, it is not self-consistent with the equilibrium. The second case (\texttt{QA\_Case2}) is a QA configuration with $n_0=n_{e,0}=2.38 \times 10^{20} $ m$^{-3}$, $T_0=T_{e,0}=9.45$ keV, that was optimized for volume averaged $\beta=$2.5$\%$ (see Sect. VIC in \cite{landreman2022optimization}). Here, $\beta$ is the ratio of the plasma pressure to the magnetic pressure.  Fig.~\ref{fig:2} and \ref{fig:3} provide further details of each configuration including the equilibrium profiles, cross-sections at various toroidal angles, and three-dimensional views.

\begin{figure}
  \centering
  \includegraphics{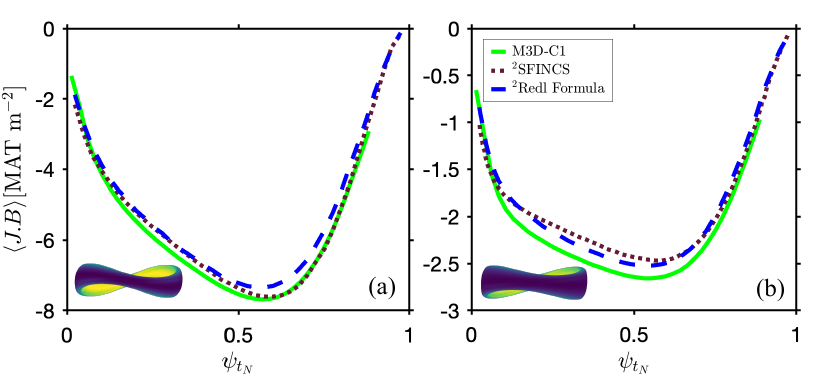}
  \caption{Bootstrap current profiles for (a) quasi-axissymetric configuration (\texttt{QA\_Case1}) and (b) optimized quasi-axisymmetric configuration with volume averaged $\beta=$2.5$\%$.(\texttt{QA\_Case2}) $^2$ \cite{landreman2014comparison}}
\label{fig:4}
\end{figure}

Figure \ref{fig:4} compares M3D-C1's bootstrap current profile with those of SFINCS and Redl formulae from \cite{landreman2022optimization}. It is evident from Fig. \ref{fig:4} that M3D-C1's implementation of the modified \cite{redl2021new} formulae as presented in \cite{landreman2022optimization} is in close agreement with the SFINCS and Redl based on calculations from \cite{landreman2022optimization}. Minor discrepancies between the profiles may be attributed to numerical treatments between the codes, although the exact sources remain unclear. Nevertheless, the results strongly support the robustness of M3D-C1's implementation in these complex configurations.

\subsection{Stellarator: Nonlinear evolution}
In this Section, the impact of the bootstrap model on the nonlinear evolution of a QA stellarator equilibrium is analyzed. The simulations conducted in M3D-C1 are initialized using the QA equilibrium profile from Section VIC of \cite{landreman2022optimization} (\texttt{QA\_Case2}). The nonlinear evolution of the equilibrium are begun from an initial state with nested magnetic surfaces. This equilibrium, developed using SIMSOPT optimization software (\cite{landreman2021simsopt}), was optimized for low quasi-symmetry error, good energetic particle confinement, and a self-consistent bootstrap current. The initial equilibrium profiles used in the simulations in this section are shown in Fig. \ref{fig:3}a. The computational domain is aligned with the shape of the equilibrium plasma, bounded by the last closed flux surface. The viscosity coefficient is fixed at ($\nu= 3.65\times10^{-4}$ kg/ms). The heat transport is modeled according to Eq. (\ref{eq:heat}), which accounts for thermal energy exchange and diffusion across the plasma. To explore a range of physical scenarios, several simulation configurations are examined. Four resistivity profiles are considered, using the Spitzer resistivity model with varying scaling factors of $\eta_0=$ \{1,10,1000,10000\}. The general resistivity form is expressed as:
\begin{equation}
\eta(R,\phi,Z)=\eta_{norm}\,\eta_0\,T^{-3/2}_e(R,\phi,Z).
\end{equation}
where $\eta_{norm}=2.74$ $\Omega$m is the normalization factor for the classical resistivity value, $\eta_0$ is an artificial scale factor, and $T_e$ is the electron temperature at the position $(R,\phi,Z)$. In this set of simulations, the perpendicular thermal conductivity ($\kappa$) is held constant at $10^{-6}$ m$^{-1}$s$^{-1}$. 
To assess the effect of the bootstrap model, all configurations are evaluated with the bootstrap model both enabled and disabled.

The simulations are performed on a semi-structured grid consisting of 36 toroidal planes, resulting in a total of $1.88\times 10^5$ three-dimensional elements within the computational domain. The grid is designed to provide sufficient resolution to accurately capture the non-linear dynamics of the plasma.  

\begin{figure}
  \centering
  \includegraphics{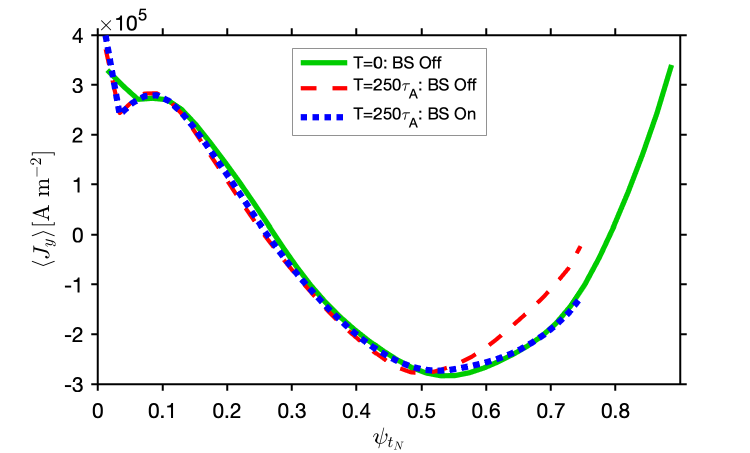}
  \caption{Toroidal current density profiles for simulations with resistivity scale factor $\eta_0=10^4$, comparing simulations with and without the bootstrap (BS) model at t=$0$ and $250\tau_A$, highlighting the effect of the bootstrap current model on the profile evolution.}
\label{fig:5}
\end{figure}

\begin{figure}
  \centering
  \includegraphics[width=\textwidth]{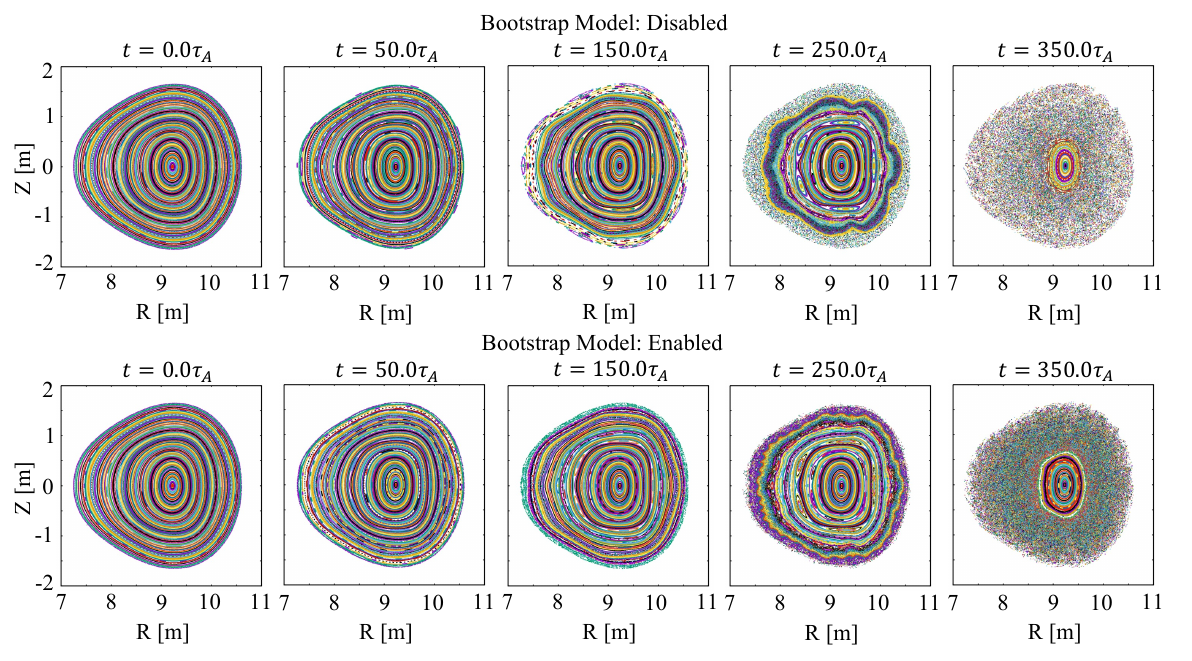}
  \caption{Poincar\'e sections of the magnetic field starting from a quasi-axisymmetric stellarator equilibrium optimized at 2.5\% plasma beta for the case with resistivity scale factor $\eta_0=10^4$ (\texttt{QA\_Case2}). The sections are shown as a function of time for $0 \leq t \leq 350\tau_A $, comparing simulations with the bootstrap model disabled (top row) and enabled (bottom row).}
\label{fig:6}
\end{figure}

Figure \ref{fig:5} shows the toroidal current density profiles from simulations with $\eta_0=10000$ at two times, $t=0$ and $t=250\tau_A$, comparing the cases with and without the bootstrap current model. In the absence of the bootstrap model, the current decays over time. In contrast, when the bootstrap model is enabled, the current is maintained, emphasizing its role in sustaining plasma currents. However, despite these differences, the simulations remain MHD unstable, and the current continues to evolve. The plasma instability is further evidenced by the Poincar\'e plots in Fig. \ref{fig:6}. At $t=250.0\tau_A$ in Fig. \ref{fig:6}, these plots reveal that the chaotic region at the plasma boundary is significantly smaller when the bootstrap model is enabled, indicating that enabling the bootstrap model modifies the evolution of instabilities by slowing the breakup of magnetic surfaces.

\begin{figure}
  \centering
  \includegraphics[width=\textwidth]{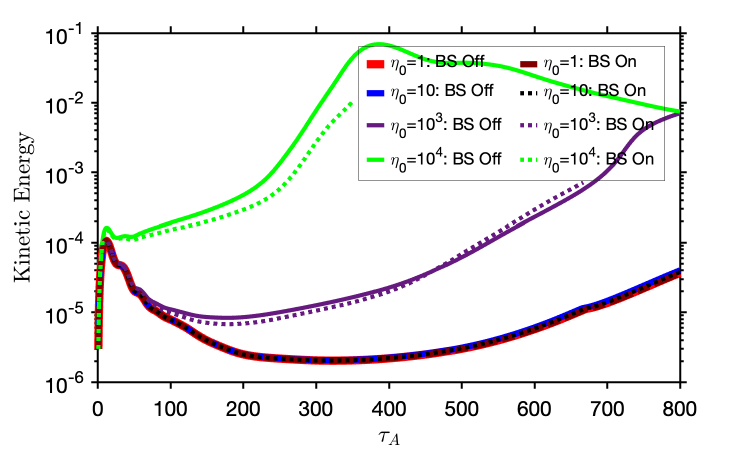}
  \caption{Time evolution of total kinetic energy for varying resistivity scaling factors $\eta_0 =$ 1, 10, 1000, and 10000 demonstrating faster kinetic energy growth at higher resistivity.}
\label{fig:7}
\end{figure}

It was found in \cite{wright2024investigating} that the growth of instabilities in equilibria similar to the ones under consideration here exist in the limit of low resistivity (\textit{e.g.}, are essentially ideal), but are accelerated at high resistivity.  Because the instabilities progressed more rapidly than the resistive decay time, it was concluded that this result was not a consequence of neglecting the bootstrap current drive; for example, through the spurious resistive decay of equilibrium currents that are actually driven by the bootstrap effect. Now that we are in a position to evaluate this with a bootstrap model included, we vary the resistivity scaling factor $\eta_0$ here, running simulations with $\eta_0=$1, 10, 1000, and 10000. Figure~\ref{fig:7} shows the kinetic energy grows more rapidly with increasing resistivity, supporting the conclusions of \cite{wright2024investigating} that resistivity enhances the growth rate, even when the bootstrap model is included.

\begin{figure}
  \centering
  \includegraphics{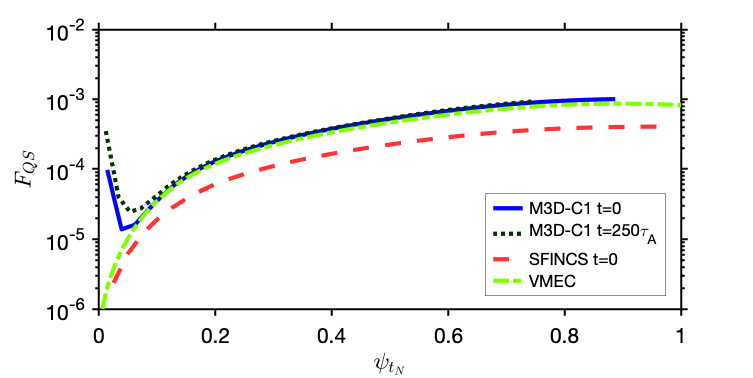}
  \caption{Two-term quasisymmetry error as defined in Eq. \ref{eq:Fqs} for the quasi-axisymmetric stellarator equilibrium optimized at 2.5\% plasma beta (\texttt{QA\_Case2}) with a resistivity scale factor $\eta_0=10^4$.}
\label{fig:8}
\end{figure}

To further investigate the stability of the equilibrium over time, the deviation from quasisymmetry is quantified as the system evolves. This is achieved by calculating the two-term quasisymmetry error ($F_{QS}$)(\cite{helander2008intrinsic,helander2014theory,paul2020adjoint}), as shown in the following equation:

\begin{equation}
F_{QS}= \left\langle \left( \frac{1}{B^3}[(N-\iota M)\vec B \times \nabla B \cdot \nabla\psi - (MG-NI)\vec B \cdot \nabla B]\right)^2 \right\rangle \label{eq:Fqs}
\end{equation}
where, $B$ is the magnetic field, $M=1,N=0$ for quasi-axisymmetry. The notation $\langle \cdot \rangle$ denotes the magnetic surface average, with isotherms of $T_e$ serving as proxies for magnetic surfaces, consistent with the approach used in the bootstrap current calculation. Fig. \ref{fig:8} compares $F_{QS}$ obtained from M3D-C1 outputs with the bootstrap model enabled, at two distinct times $t=0.0$, $t=250.0\tau_A$, as well as from SFINCS and VMEC using the $t=0$ equilibrium. The M3D-C1 $F_{QS}$ metric closely resembles the VMEC quasisymmetry error. While the equilibrium evolves over time, as observed in the Poincar\'e plots, the $F_{QS}$ does not significantly change between $0<t<250$ $\tau_A$ indicating that quasisymmetry is maintained throughout this period. However, beyond $t>250$, the chaotic region grows making it difficult to generate well-defined isothermal surfaces globally (at which point neoclassical transport is likely strongly subdominant to parallel transport along chaotic fieldlines anyway), so analysis is concluded there.

\section{Summary \& Discussion}
\label{sec:conclusions}
This work enhances the capabilities of the M3D-C1 extended-MHD code by incorporating self-consistent bootstrap current models for tokamak and quasisymmetric stellarator geometries. Two models are implemented, namely, (a) the Sauter model (\cite{sauter1999neoclassical}) and (b) the Redl model (\cite{redl2021new}). For quasisymmetric stellarators, the isomorphism outlined by \cite{landreman2022optimization} is employed.

The numerical verification presented in this work demonstrates the accuracy of these new implementations. In Sec.~\ref{sec:tokamak_verification}, the bootstrap current in a tokamak with a circular plasma boundary is tested using both bootstrap models. Comparisons between the flux-surface-averaged bootstrap current from the updated M3D-C1 code and results from established neoclassical codes, such as XGCa (\cite{hager2016gyrokinetic}), NEO (\cite{belli2008kinetic,belli2011full}), and SFINCS (\cite{landreman2014comparison}), show excellent agreement.

In Sec.~\ref{sec:stellarator_verification}, the bootstrap current is evaluated in two quasisymmetric stellarator configurations, \texttt{QA\_Case1} and \texttt{QA\_Case2}. These results are compared with SFINCS calculations, yielding good agreement. Finally, Sec. 3.3 explores the application of the updated M3D-C1 code to non-linear MHD simulations of an optimized QA stellarator configuration, where the expected toroidal current sustainment is observed.

The accurate modeling of bootstrap currents is critical to understanding and improving plasma performance in magnetic confinement devices. These results not only verify the bootstrap models in the updated M3D-C1 code but also highlight its potential to improve the design and optimization of future fusion reactors through accurate neoclassical current predictions in non-axisymmetric geometries. The successful implementation and verification of these models provides a solid foundation for advancing non-linear MHD simulations, offering crucial insights that can guide the design and operational strategies of next-generation fusion devices.

This capability will enable new investigations into nonlinear MHD physics where bootstrap current is expected to be important.  In particular, this includes transport-timescale simulations where the global equilibrium evolves due to heating, as is done in nonlinear calculations of pressure limits (\textit{c.f.}, \cite{Wright_2024}), sawtooth cycles (\textit{c.f.}, \cite{Jardin_2012}), and ELM cycles (\textit{c.f.}, \cite{Futatani_2021}), for example.  While the implementation described here is only strictly valid for quasisymmetric magnetic geometries, the ability to calculate the neoclassical transport of generic nonaxisymmetric output using 3D NEO~\cite{Sinha_2022,Sinha_2023}) and SFINCS (as described above) enable \textit{post hoc} evaluation of the accuracy of the Sauter-Redl-Landreman model.  These capabilities also lay the groundwork for coupled M3D-C1/3D NEO or M3D-C1/SFINCS modeling, which would be applicable to a wider range of geometries.  One fundamental challenge is that the treatment of stochastic regions is beyond the scope of existing neoclassical theory, and requires a kinetic-MHD theory to treat self-consistently.  In cases where the accuracy of the dynamics of these regions is less critical, such as when the time evolution is slow or time-dependent effects are negligible, the flattening of temperature and density profiles by classical transport along stochastic field lines, which is naturally included in the extended-MHD model, is expected to dominate over neoclassical effects in any case.  We also note that the Sauter-Redl-Landreman models implemented here, as well as NEO 3D and SFINCS, all calculate time-independent transport, and are therefore not able to treat time-dependent neoclassical response.  In principle this limits the applicability of the model on Alfv\'enic timescales (\textit{e.g.}, kink modes)---compounded by the fact that these modes are not likely to maintain quasisymmetry---but the neoclassical response is not expected to play a significant role in such phenomena anyway.

This work was supported by the U.S. Department of Energy under contract number DE-AC02-09CH11466. The United States Government retains a non-exclusive, paid-up, irrevocable, world-wide license to publish or reproduce the published form of this manuscript, or allow others to do so, for United States Government purposes.

\appendix

\section{Bootstrap Current Formulation}\label{appA}
The bootstrap current formulation used in this study is follows Eq. \ref{eq:A1} (\cite{sauter1999neoclassical}; \cite{10.1063/1.1517052}) and Eq.\ref{eq:A2} (\cite{redl2021new}). 

\begin{equation}
\langle \boldsymbol{J_x \cdot B}\rangle=\langle j_{\parallel}B\rangle =-F(\psi)\left({pL_{31}\frac{\partial \:\text{ln}\:p}%
  {\partial \psi}+p_eL_{32}\frac{\partial \:\text{ln}\:T_e}%
  {\partial \psi}+p_i\alpha L_{34}\frac{\partial\: \text{ln}\:T_i}%
  {\partial \psi}}\right) \label{eq:A1}
\end{equation}

\begin{equation}
\langle \boldsymbol{J_x \cdot B}\rangle=\langle j_{\parallel}B\rangle =-F(\psi)\left({pL_{31}\frac{\partial\: \text{ln}\:n}%
  {\partial \psi}+p_e(L_{31}+L_{32})\frac{\partial\: \text{ln}\:T_e}%
  {\partial \psi}+p_i(L_{31}+\alpha L_{34})\frac{\partial\: \text{ln}\:T_i}%
  {\partial \psi}}\right) \label{eq:A2}
\end{equation}

where $F(\psi)=RB_\phi$, $\psi$ is poloidal flux per radian i.e. $\Psi_p=2\pi\psi$, $R$ is the major radius and $B_\phi$ the toroidal magnetic field. For details on the definitions of the coefficients in these equations, please refer to the corresponding references.

For convenience, the common terms used to calculate the coefficients in Eq.\ref{eq:A1} and Eq.\ref{eq:A2} and their definitions, as presented in \cite{sauter1999neoclassical} and \cite{redl2021new} are summarized below:

The trapped particle fraction, denoted by $f_t$ is
\begin{equation}
f_{t}=1-\frac{3}{4}\langle B^2\rangle\int_0^{1/B_{max}}\frac{\lambda d\lambda}{\langle \sqrt{1-\lambda B}\rangle}.
\end{equation}
The effective electron and ion collisionalities are
\begin{subequations}
    \begin{equation}
    \nu_e^*=6.921 \times 10^{-18} \frac{qRn_e \:\text{ln} \:\Lambda_e}{T_e^2\varepsilon^{3/2}}, \label{eq:C2a}
    \end{equation}
    \begin{equation}
    \nu_i^*=4.9 \times 10^{-18} \frac{qRZ^4n_e \:\text{ln} \:\Lambda_{ii}}{T_i^2\varepsilon^{3/2}}, \label{eq:C2b}
    \end{equation}
\end{subequations}
The Coulomb logarithms are:
\begin{subequations}
    \begin{equation}
    \Lambda_e=31.3-\text{ln}\left( \frac{\sqrt{n_e}}{T_e} \right). 
    \end{equation}
    \begin{equation}
    \Lambda_{ii}=30-\text{ln}\left( \frac{Z^3\sqrt{n_i}}{T_i^{3/2}} \right). 
    \end{equation}
\end{subequations}
Here, $R$ [m] is the major radius, $q=1/i$ is the safety factor, electron ($T_e$) and ion ($T_i$) temperatures in eV and densities ($n_e$ and $n_i$) in m$^{-3}$.

\section{Isomorphism}\label{appB}
For applicability to quasisymmetric stellarators, the \cite{redl2021new} formula is modified according to the isomorphism as defined in \citet[]{landreman2022optimization}, see Eq. \ref{eq:Landreman}. 
 
Noting that the rotational transform is defined as $\partial \psi/\partial \psi_t = \iota$, we have 
\[
\frac{\partial}{\partial \psi_t} = \iota \frac{\partial}{\partial \psi}
\]
In case of axisymmetry, $N=0$ and $G=I=RB_\phi$ in Eq. \ref{eq:Landreman} recovers the result of \cite{redl2021new}. For the calculation of the bootstrap coefficients, the factors $qR$ in equations \ref{eq:C2a} and \ref{eq:C2b} and the inverse aspect ratio $\epsilon$ are defined as follows:
\begin{subequations}
    \begin{equation}
    qR=\frac{G+iI}{i-N} \left\langle {\frac{1}{B}}\right \rangle,
    \end{equation}
    \begin{equation}
    \epsilon=\frac{B_{max}-B_{min}}{B_{max}+B_{min}}.
    \end{equation}
\end{subequations}

\section{M3D-C1: Single Fluid Equations}\label{appC}
Summary of the M3D-C1 equations used in this study (reproduced from \cite{jardin2012multiple}). 
\begin{equation}
\frac{\partial n}{\partial t}+\boldsymbol{\nabla} \cdot (n \boldsymbol{v})=0 ,\:\:\:\:\text{continuity},
\end{equation}
\begin{equation}
nm_i\left(\frac{\partial \boldsymbol{v}}{\partial t}+\boldsymbol{v \cdot \nabla v} \right)=\boldsymbol{J} \times \boldsymbol{B}-\nabla p-\boldsymbol{\nabla \cdot \Pi}+\boldsymbol{{F}}, \:\:\:\:\text{momentum},\label{eq:momentum}
\end{equation}
\begin{equation}
    \begin{aligned}
        \frac{\partial p}{\partial t}+\boldsymbol{v} \cdot \nabla p+ \Gamma p\nabla \cdot \boldsymbol{v} &=(\Gamma-1)[\eta J^2-\boldsymbol{\nabla \cdot q} - \boldsymbol{\Pi:\nabla v} +Q],\:\:\:\:\text{energy},\\
        &=(\Gamma-1)[\eta \boldsymbol{J \cdot (J-J_x)}-\boldsymbol{\nabla \cdot q} - \Pi:\nabla \boldsymbol{v} +Q]\label{eq:modifiedenergy}
    \end{aligned}
\end{equation}

\begin{subequations}
    \begin{equation}
        \begin{aligned}
        \frac{\partial \boldsymbol{B} }{\partial t}&=\boldsymbol{\nabla \times (v \times B}-\eta [\boldsymbol{J}])\\
        &=\boldsymbol{\nabla \times (v \times B}-\eta [\boldsymbol{J}-\boldsymbol{J}_x]),\label{eq:modifiedohms}
        \end{aligned}
    \end{equation}
    \begin{equation}
    \boldsymbol{J}=\frac{1}{\mu_0}\boldsymbol{\nabla \times B},\:\:\:\:\text{Maxwell},
    \end{equation}
\end{subequations}

    \begin{equation}
        \begin{aligned}
        \boldsymbol{q}=\kappa_\perp \nabla T_e -\kappa_{\parallel} \frac{\boldsymbol{B}\boldsymbol{B}}{B^2} \cdot \nabla T_e,\label{eq:heat}\:\:\:\:\text{heat transport model}
        \end{aligned}
    \end{equation}

where $n$ is the density, $\boldsymbol{v}$ is the fluid velocity, $m_i$ is ion mass, $\boldsymbol{J}$ is the current density, $\boldsymbol{B}$ is the magnetic field, $p$ is the pressure, $\Pi$ is the viscous stress tensor, $\boldsymbol{F}$ is the external force, $Q$ is external heat source, $\Gamma =$ 5/3, $\eta$ is the resistivity, $\boldsymbol{q}$ is the heat flux, ${\mu_0}$ is the vacuum permeability.

\bibliographystyle{jpp}

\bibliography{jpp_new}

\end{document}